# $CO_2$-Activation and Enhanced Capture by $C_6Li_6$: A Density Functional Approach

Ambrish Kumar Srivastava

Department of Physics, DDU Gorakhpur University, Gorakhpur-273009, Uttar Pradesh, India

Corresponding author

E-mail: ambrishphysics@gmail.com; aks.ddugu@gmail.com

Tel. No.: +91 9415620016

**Abstract**

The capture and storage of $CO_2$, a major component of greenhouse gases, are crucial steps that can positively impact the global carbon balance. The capture of $CO_2$ has been difficult due to its extremely high stability. In this study, we propose a simple and yet effective approach for capture and storage of $CO_2$ by $C_6Li_6$. $C_6Li_6$ possesses a planar star-like structure, whose ionization energy is lower than that of Li atom and hence, it behaves as a superalkali. Superalkalis are unusual species possessing lower ionization energies than alkali atoms. We have systematically studied the interaction of successive $CO_2$ molecules with $C_6Li_6$ using long-range dispersion corrected density functional ωB97xD/6-311+G(d) calculations. We notice that these interactions lead to stable $C_6Li_6$-$nCO_2$ complexes ($n$ = 1-6) in which the structure of $CO_2$ moieties is bent appreciably (122-125°) due to electron transfer from $C_6Li_6$, whose planarity is distorted only slightly (≤ 7°). This clearly suggests that the $CO_2$ molecules can successfully be activated and captured by $C_6Li_6$. We have also analyzed bond-lengths and bond-angle of $CO_2$, their charges and adsorption energy as a function of the number of adsorbed $CO_2$ ($n$). It has been also noticed that the bond-length of $CO_2$ in $C_6Li_6$-$nCO_2$ complexes increases monotonically whereas adsorption energy decreases, ranging 3.18-2.79 eV per $CO_2$ with the increase in $n$. These findings not only establish the potential of $C_6Li_6$ for capture and storage of $CO_2$ molecules but also provide new insights into $CO_2$-activation, capture, and storage by systems having low ionization energies.

## 1. Introduction

The capture and storage of $CO_2$, a major component of greenhouse gases, are crucial steps that can positively impact the global carbon balance [1, 2]. To address these important issues, a number of methods have been devised to sequestrate [3] and incorporate $CO_2$ into other molecules [4, 5]. The major challenge in the recycling of $CO_2$ into usable fuels is that $CO_2$ is an extremely stable molecule. To convert $CO_2$ into fuel, it should be first activated or chemically reduced by some catalysts. Due to the negative electron affinity of $CO_2$ [6], the single-electron reduction of $CO_2$ has not been easy. Recently, however, it has been suggested [7-9] that a special class of species can be employed to reduce $CO_2$ into $CO_2^-$. These species possess lower ionization energy (IE) than those of alkali metal and therefore, are referred to as superalkalis [10]. Superalkalis have been previously employed in the design of supersalts [11-13], superbases [14, 15], alkalides [16-18], etc. Due to their ever-increasing applications, superalkalis have been subject of continuous investigations to date.

Recently, Zhao *et al.* [7] have proposed a rational design of superalkalis and their role in $CO_2$-activation. They have employed novel superalkalis to reduce $CO_2$ into $CO_2^-$. Subsequently, the capability of reducing $CO_2$ by typical superalkalis is verified by Srivastava [8], Park and Meloni [9]. It was noticed that superalkalis can indeed be used to reduce $CO_2$ into $CO_2^-$ anion. This might open an opportunity to use superalkalis in the capture and storage of $CO_2$. However, activation of $CO_2$ is limited to one molecule per superalkali system, i.e., a superalkali can reduce only one $CO_2$ molecule as reported by previous works [7-9]. Therefore, the $CO_2$-storage capacity of these superalkalis seems too poor to be used for practical applications.

Herein, we report the application of $C_6Li_6$ for activation and sequential capture of $CO_2$. We demonstrate that $C_6Li_6$ is capable to reduce and capture up to six $CO_2$ molecules sequentially. Note that $C_6Li_6$ has been previously studied by several groups [19-22] and its application in the hydrogen storage has also been reported [23, 24]. The low IE feature of C6Li6 was particularly highlighted only in a recent study [25]. It was noticed [25] that the IE of $C_6Li_6$ is lower than that of Li atom and consequently, it may be treated as a superalkali. Unlike other superalkalis, however, it possesses a closed-shell structure. Therefore, it seems interesting to study the $CO_2$-reduction by $C_6Li_6$. We shall see that $C_6Li_6$ is not only capable of capturing $CO_2$ molecules but also effective in their storage.

## 2. Computational details

All computations were performed using the density functional theory (DFT) approach at ωB97xD level [26] and 6-311+G(d) basis set in Gaussian 09 program [27]. The total energy and geometry have been obtained without any symmetry constraints in the potential energy surface. The vibrational analysis has been performed at the same level of theory to ensure that all frequency values are positive, i.e., the optimized structures belong to true minima in the potential energy surface. The partial charges have been computed by natural bond orbital (NBO) analysis [28] as implemented in Gaussian 09. Although this program offers a variety of functionals, the ωB97xD functional incorporates long-range and dispersion corrections implicitly.

Li *et al.* [29] studied a series of complexes consisting of $CO_2$ and amines and verified the performance of various density functionals. They noticed that the long-range corrected functionals (with RMS errors of 0.5 kcal/mol) were preferred if the corresponding DFT optimized geometries were adapted for the benchmark. In order to further assess the performance of ωB97xD/6-311+G(d) scheme, we have performed some test calculations on $C_6Li_6$ and $CO_2$ as listed in Table 1. We noticed that our computed bond lengths of $C_6Li_6$ and $CO_2$ are in good agreement with the corresponding literature values [25, 30]. Furthermore, the ionization energy of $C_6Li_6$ and electron affinity of $CO_2$ are also in accordance with their reported values [6, 25].

This suggests the validity of present computational scheme and reliability of the calculated results.

### 3. Results and discussion

#### 3.1. Equilibrium structures of $C_6Li_6$-$nCO_2$ complexes

$C_6Li_6$ possesses a planar structure having ring bond lengths of 1.418 Å whereas neutral $CO_2$ is a linear molecule with the bond length of 1.157 Å as displayed in Fig. 1. We have also included the structure of $CO_2^-$, which is bent by an angle of 136.9° with the bond length of 1.229 Å. Therefore, single-electron reduction of $CO_2$ activates it by bending and increasing the bond length. We have studied the interaction of $CO_2$ with $C_6Li_6$, which leads to the formation of complexes *viz.* $C_6Li_6$-$nCO_2$ as depicted in Fig. 2. In all complexes, the interaction takes place *via* one C-C and two Li-O bonds between $CO_2$ and $C_6Li_6$. The structural parameters associated with $C_6Li_6$-$nCO_2$ complexes such as bond-lengths of $C_6Li_6$, bond- angle of $CO_2$ as well as interaction bond lengths between $CO_2$ and $C_6Li_6$ are listed in Table 2. The interaction of $CO_2$ leads to displace Li atoms out-of-plane of the $C_6Li_6$ ring. In $C_6Li_6$-$CO_2$, $C_6Li_6$ ring, having bond-lengths in the range 1.412-1.423 Å, is distorted by 7°. The bond-lengths and angle of $CO_2$ are 1.263 Å and 122.5°, respectively whereas, the interaction bond-lengths $R_{Li-O}$ and $R_{C-C}$ are 1.775 and 1.540 Å, respectively.

$C_6Li_6$-$2CO_2$ possesses an almost similar structure as that of $C_6Li_6$-$CO_2$, although the ring-distortion ($\delta_{ring}$) is significantly reduced to 0.3°. This is evidently due to two $CO_2$ binding symmetrically to the $C_6Li_6$ ring. In $C_6Li_6$-$3CO_2$, on the contrary, the ring is perfectly planar like $C_6Li_6$ having equal bond-lengths of 1.418 Å. The $R_{Li-O}$ and $R_{C-C}$ in this complex are slightly larger and smaller than those in $C_6Li_6$-$CO_2$ and $C_6Li_6$-$2CO_2$, respectively. Although the bond-angle of $CO_2$ moieties is slightly increased to 123.9°, the $CO_2$ bond-lengths remain unaltered as compared to those in $C_6Li_6$-$CO_2$ and $C_6Li_6$-$2CO_2$. In both $C_6Li_6$-$4CO_2$ and $C_6Li_6$-$5CO_2$, $C_6Li_6$ rings possess bond-lengths in the range 1.407-1.412 Å and 1.403-1.408 Å, respectively. The bond lengths and angle of $CO_2$ moieties range 1.251-1.261 Å and 123.9-125.3°, respectively. Likewise, $R_{Li-O}$ and $R_{C-C}$ in these complexes are 1.781-1.807 Å and 1.519-1.535 Å, respectively. The $C_6Li_6$ ring in $C_6Li_6$-$6CO_2$ is perfectly planar, like $C_6Li_6$-$3CO_2$, having equal bond-lengths of 1.400 Å. The bond-length and angle of $CO_2$ moieties are 1.251 Å and 125.2°, respectively whereas the interaction bond-lengths $R_{Li-O}$ and $R_{C-C}$ are 1.799 Å and 1.529 Å, respectively.

To get some insights into the electronic structure of $C_6Li_6$-$nCO_2$ complexes, we have analyzed their highest occupied molecular orbitals (HOMOs). The HOMOs of $C_6Li_6$-$nCO_2$ complexes are plotted in Fig. 3. For comparison, the HOMO of $C_6Li_6$ is also displayed. One can note that the HOMO of $C_6Li_6$-$nCO_2$ complexes ($n$ = 1-3) are mainly localized on the $C_6Li_6$ ring. These HOMOs resemble that of $C_6Li_6$, with a little or no contribution of $CO_2$. For $n > 3$, however, the contribution of $CO_2$ in the HOMO increases. The HOMO of $C_6Li_6$-$6CO_2$, for instance, is significantly contributed by $CO_2$ moieties (see Fig. 3).

### 3.2. Activation and sequential adsorption of $CO_2$ by $C_6Li_6$

As mentioned earlier, the activation of $CO_2$ leads to an increase in the bond-length and decrease in the bond-angle by bending. In Fig. 4, we have plotted the average bond-length and bond-angle of $CO_2$ moieties in $C_6Li_6$-$nCO_2$ complexes. In the case of $C_6Li_6$-$CO_2$, the bond-length and angle of $CO_2$ 1.263 Å and 122.5° are slightly larger and smaller than the corresponding values of $CO_2^-$. The bond-lengths and angle of $CO_2$ in superalkali-$CO_2$ complexes have been reported as 1.246 Å and 133°, respectively for $FLi_2$, $OLi_3$ and $NLi_4$ superalkalis [8] and 1.21-1.29 Å and 131-137°, respectively for $F_2Li_3$ superalkali [9].

As the number of $CO_2$ increases, the average bond-length decreases monotonically and reaches 1.251 Å for $C_6Li_6$-$6CO_2$. On the contrary, the average bond-angle of $CO_2$ increases with the increase in the number of $CO_2$ and becomes 125.2° for $C_6Li_6$-$6CO_2$. Since $CO_2$ molecules are captured by the charge transfer from $C_6Li_6$, we have calculated the partial natural charge ($\Delta q$) located at $CO_2$ moieties and listed in Table 3 (also plotted in Fig. 4). One can note that the $\Delta q$ of $C_6Li_6$-$nCO_2$ is approximately -0.83$e$ for $n$ = 1-3, -0.81$e$ for $n$ = 4, -0.79$e$ for $n$ = 5 and -0.77$e$ for $n$ = 6. Note that the $\Delta q$ values have been found to be -0.90$e$ for $FLi_2$-$CO_2$, -0.88$e$ for $OLi_3$-$CO_2$ and -0.85$e$ for $NLi_4$-$CO_2$ complexes [8]. Similarly, the $\Delta q$ value reported for the most stable structure of the $F_2Li_3$-$CO_2$ complex is -0.78$e$ [9]. These $\Delta q$ values clearly suggest that the adsorbed $CO_2$ molecules are reduced to $CO_2^-$. Thus, $C_6Li_6$ can be successfully used for the single-electron reduction, i.e., activation of all $CO_2$ molecules and consequently, their adsorption.

The adsorption energy ($E_{ad}$) per $CO_2$ molecule and consecutive adsorption energy ($\Delta E_{ad}$) are calculated by using the following equations;

$E_{ad} = [E(C_6Li_6) + n\, E(CO_2) - E(C_6Li_6\text{-}nCO_2)]/n$

$\Delta E_{ad} = E[C_6Li_6\text{-}(n\text{-}1)CO_2] + E[CO_2] - E[C_6Li_6\text{-}nCO_2]$

where $E[..]$ represents total (electronic + zero-point) energy of respective species and $n$ = 1-6. The calculated $E_{ad}$ values are listed in Table 3 and also plotted in Fig. 4. One can note that the $E_{ad}$ values are positive, ranging 3.18-2.79 eV per $CO_2$. The $E_{ad}$ value of $C_6Li_6$-$CO_2$ (3.18 eV) is larger than the binding energy value of superalkali-$CO_2$ complexes reported in previous works [8, 9]. This may suggest that all $C_6Li_6$-$nCO_2$ complexes are stable. More interestingly, the highest $E_{ad}$ of 3.31 eV corresponds to $C_6Li_6$-$3CO_2$ (see Fig. 4) in which three $CO_2$ are adsorbed at alternate positions leaving a perfectly planar ring structure as mentioned earlier. The $\Delta E_{ad}$ is the energy gained by successive adsorption of $CO_2$ molecules and an important index for testing the continuous (sequential) adsorption capacity of materials. The adsorption of molecules is difficult if the $\Delta E_{ad}$ is negative whereas the positive $\Delta E_{ad}$ values suggest that the sequential adsorption can occur between the $CO_2$ molecule and the $C_6Li_6$. Both $E_{ad}$ and $\Delta E_{ad}$ values clearly reveal the potential of $C_6Li_6$ to reduce and capture six $CO_2$ molecules successively. In order to check whether temperature and entropy effects may destabilize $C_6Li_6$-$nCO_2$ complexes, we have analyzed the Gibbs' free energy change ($\Delta G_{ad}$) during $CO_2$-adsorption calculated as below:

$$\Delta G_{ad} = \Delta G[C_6Li_6\text{-}nCO_2] - \Delta G[C_6Li_6] - n\,\Delta G[CO_2]$$

The $\Delta G$ values of respective species are computed at the room-temperature. From Table 3, one can see that all $\Delta G_{ad}$ values are negative. Furthermore, the magnitude of $\Delta G_{ad}$ increases with the increase in the number of $CO_2$ molecules adsorbed by $C_6Li_6$. These values suggest that $C_6Li_6$-$nCO_2$ complexes are thermodynamically stable as the $CO_2$-adsorption process is exothermic and spontaneous at the room-temperature.

## 4. Conclusions

A systematic study on the $CO_2$ activation and capture by $C_6Li_6$ has been performed using ωB97xD/6-311+G(d) calculations. We have noticed that $C_6Li_6$ is not only capable of reduction but also effective in the sequential adsorption of six $CO_2$ molecules. We have obtained the equilibrium structures of $C_6Li_6$-$nCO_2$ complexes ($n$ = 1-6) and noticed that planar $C_6Li_6$ ring is distorted only slightly ($\leq 7^\circ$) by $CO_2$-adsorption. In all these complexes, the partial charge on $CO_2$ moieties ranges between -0.77$e$ and -0.83$e$, suggesting the single-electron reduction of $CO_2$ molecules. These complexes are electronically and thermodynamically stable due to positive adsorption energy and negative Gibbs' free energy change during $CO_2$-adsorption. We have also analyzed the properties of $C_6Li_6$-$nCO_2$ complexes as a function of the number of $CO_2$ ($n$). We have noticed that the bond-length of $CO_2$ increases whereas bond angle decreases as *n* increases in these complexes. Similarly, the adsorption energy decreases with the increase in *n*, ranging 3.18-2.79 eV per $CO_2$. Thus, $C_6Li_6$ is indeed capable in the capture and storage of $CO_2$ molecules.

**Acknowledgement**

Dr. A. K. Srivastava acknowledges Prof. N. Misra, Department of Physics, University of Lucknow, for providing computational support and Prof. S. N. Tiwari, Department of Physics, DDU Gorakhpur University, for helpful discussions.

Table 1. Test calculations performed over $C_6Li_6$ and $CO_2$ at $\omega$B97xD/6-311+G(d) level.

| Parameter | $C_6Li_6$ | | $CO_2$ | |
|---|---|---|---|---|
| | This work | Literature | This work | Literature |
| Bond length (Å) | 1.418 | 1.422[a], 1.420[b] | 1.157 | 1.160[c] |
| Ionization energy (eV) | 4.42 | 4.48[b] | - | - |
| Electron affinity (eV) | - | - | -0.55 | -0.60[d] |

a) Ref. [19]
b) Ref. [25]
c) Ref. [30]
e) Ref. [6]

Table 2. Structural parameters of $C_6Li_6$-$n$$CO_2$ complexes at ωB97xD/6-311+G(d) level for various $n$ values.

| $n$ | $C_6Li_6$ ring | | $CO_2$ moiety | | Interaction lengths | |
|---|---|---|---|---|---|---|
| | Bond-length (Å) | $\delta_{ring}$ (°) | Bond-length (Å) | Angle (°) | $R_{Li-O}$ (Å) | $R_{C-C}$ (Å) |
| 1 | 1.412-1.423 | 7.0 | 1.263 | 122.5 | 1.775 | 1.540 |
| 2 | 1.418-1.425 | 0.3 | 1.262 | 122.7 | 1.776 | 1.540 |
| 3 | 1.416 | 0 | 1.261 | 123.9 | 1.805 | 1.535 |
| 4 | 1.407-1.412 | 3.5 | 1.261-1.259 | 123.9-124.9 | 1.782-1.806 | 1.535-1.519 |
| 5 | 1.403-1.408 | 1.7 | 1.251-1.259 | 123.9-125.3 | 1.781-1.807 | 1.522-1.530 |
| 6 | 1.400 | 0 | 1.251 | 125.2 | 1.799 | 1.529 |



Table 3. The partial NBO charge on $CO_2$ ($\Delta q$), adsorption energy per $CO_2$ ($E_{ad}$), consecutive adsorption energy ($\Delta E_{ad}$) and Gibbs' free energy change ($\Delta G_{ad}$) during $CO_2$-adsorption for $C_6Li_6$-$nCO_2$ complexes obtained at ωB97xD/6-311+G(d) level for various $n$ values.

| $n$ | $\Delta q$ ($e$) | $E_{ad}$ (eV) | $\Delta E_{ad}$ (eV) | $\Delta G_{ad}$ (kcal/mol) |
|---|---|---|---|---|
| 1 | -0.829 | 3.18 | 3.18 | -60.4 |
| 2 | -0.827 | 3.16 | 3.11 | -138.6 |
| 3 | -0.826 | 3.31 | 3.63 | -189.3 |
| 4 | -0.807[a] | 3.07 | 2.33 | -231.7 |
| 5 | -0.790[a] | 2.91 | 2.27 | -271.3 |
| 6 | -0.770 | 2.79 | 2.21 | -310.2 |



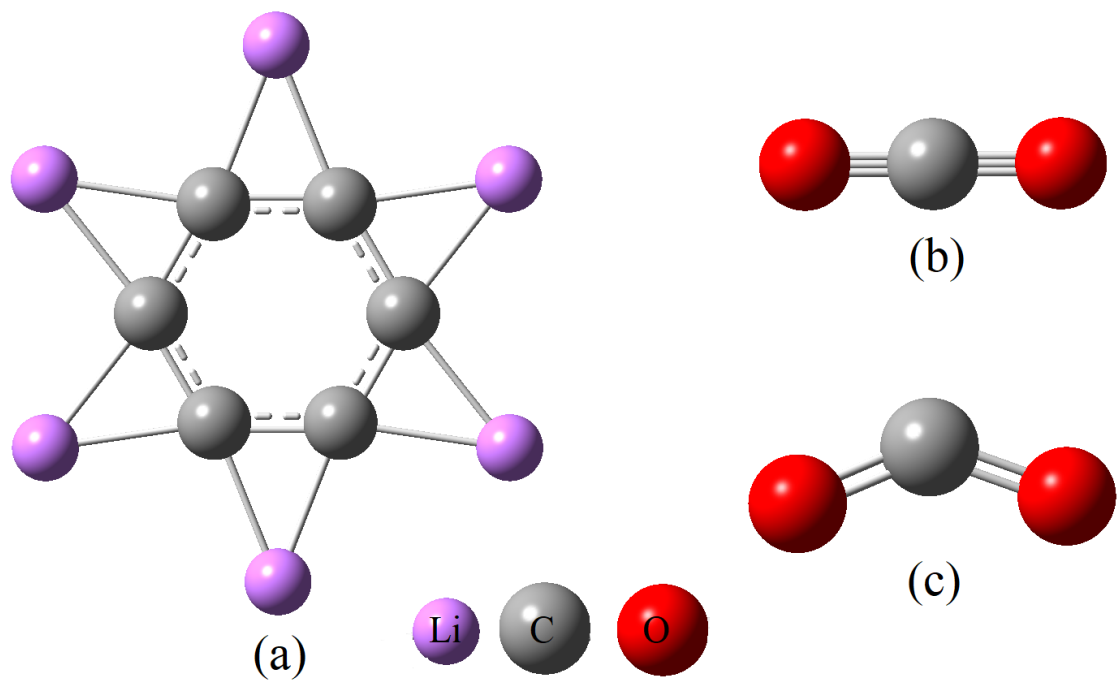

Fig. 1. Equilibrium structures of $C_6Li_6$ (a), $CO_2$ molecule (b) and $CO_2^-$ anion (c) obtained at ωB97xD/6-311+G(d) level.



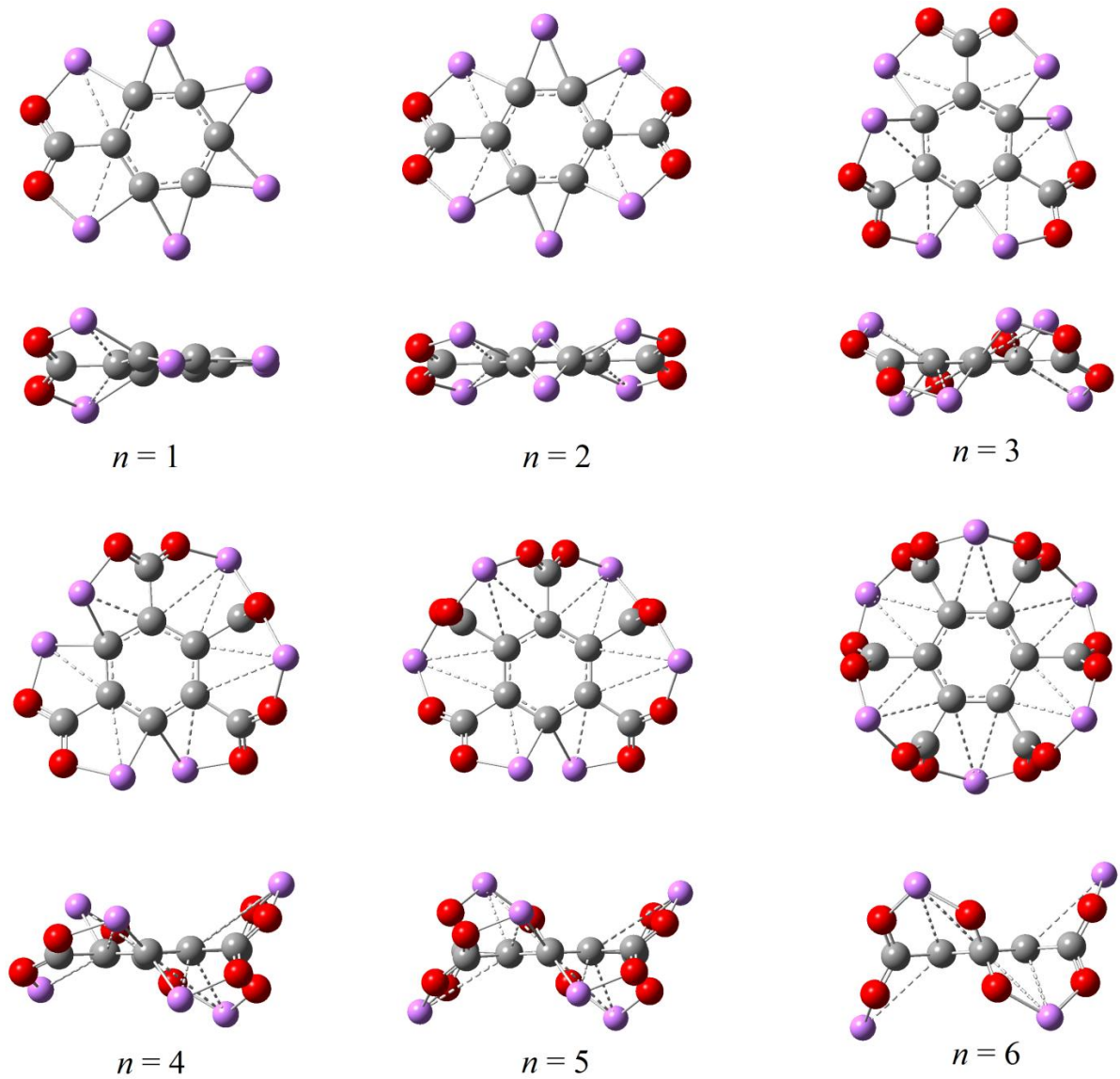

Fig. 2. Equilibrium structure of $C_6Li_6$-$n$$CO_2$ complexes for $n$ = 1-6 obtained at ωB97xD/6-311+G(d) level. Front as well as side views are also displayed.

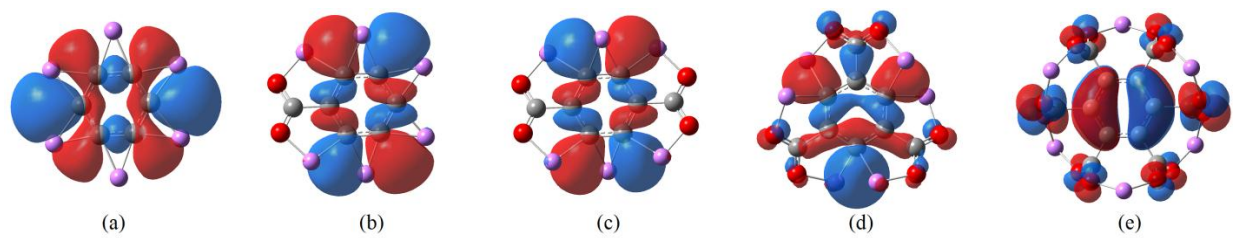

Fig. 3. The highest occupied molecular orbitals of $C_6Li_6$-$nCO_2$ complexes ($n$ = 1, 2, 3 and 6) along with that of $C_6Li_6$ with an isovalue of 0.02 a.u.

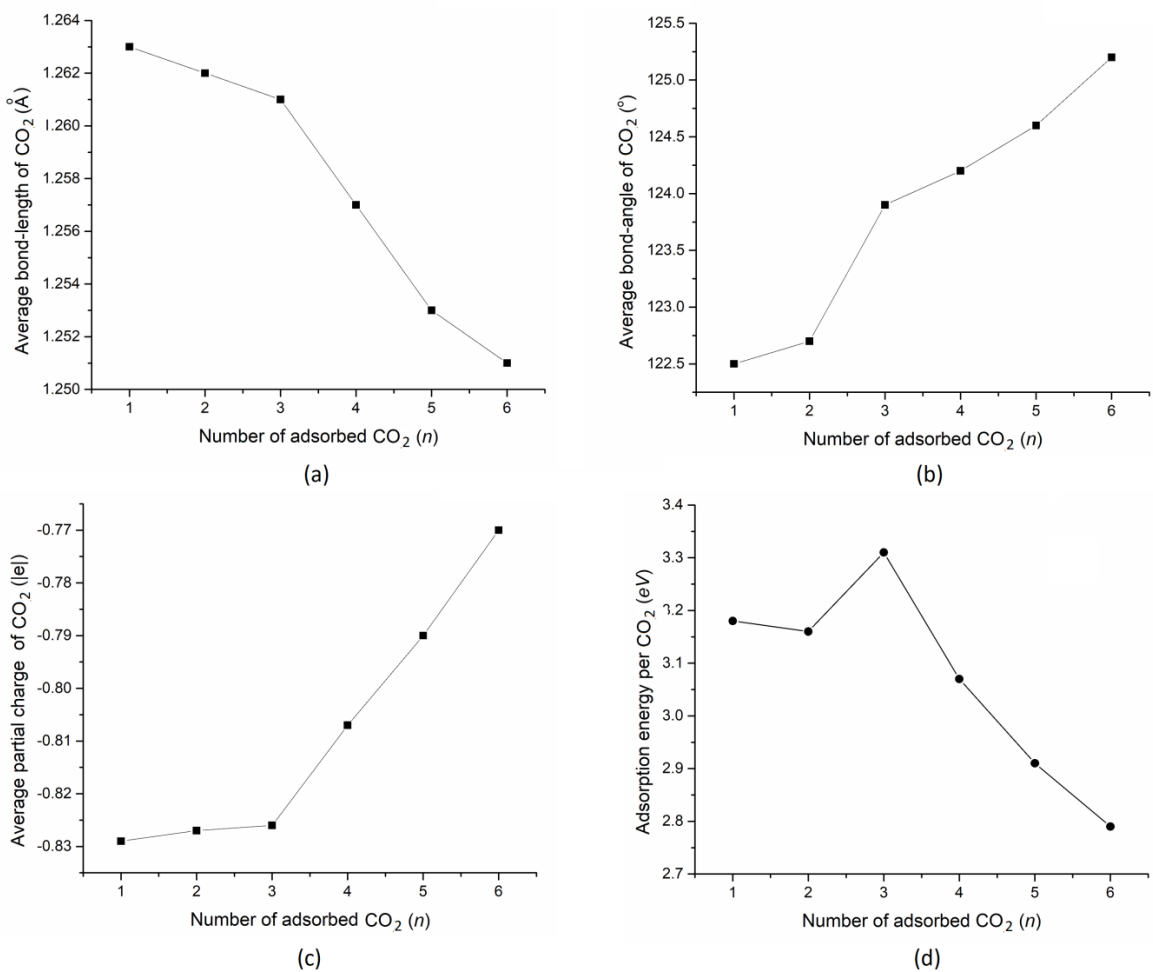

Fig. 4. The bond-lengths (a) and bond-angle (b), average partial charge (c) of adsorbed $CO_2$ and adsorption energy per $CO_2$ (d) plotted as a function of the number of $CO_2$ ($n$) in $C_6Li_6$-$n CO_2$ complexes.